\documentstyle[12pt,twoside,fleqn,espcrc1,epsf]{article}
\newcommand{\AmS}{{\protect\the\textfont2
  A\kern-.1667em\lower.5ex\hbox{M}\kern-.125emS}}

\title{Signatures of Absorption Mechanisms for $J/\psi$ and $\psi'$
Production in High-Energy Heavy-Ion Collisions}

\author{Cheuk-Yin Wong\address{Physics Division, Oak Ridge National
Laboratory, Oak Ridge, TN 37831 }%
\thanks{ This research was supported by the
Division of Nuclear Physics, U.S. D.O.E.  under Contract
DE-AC05-96OR22464 managed by Lockheed Martin Energy Research Corp.}
}

\begin{document}
\maketitle

\begin{abstract}
$J/\psi$ and $\psi'$ produced in high-energy heavy-ion collisions are
absorbed by their collisions with nucleons and produced soft
particles, leading to two distinct absorption mechanisms.  The
signature of absorption by produced soft particles, as revealed by
$\psi'$ production data, consists of a gap and a change of the slope
in going from the $pA$ line to the nucleus-nucleus line when we make a
semi-log plot of the survival probability as a function of the path
length.  Using this signature, we find from the $J/\psi$ production
data in $pA$, O-Cu, O-U, and S-U collisions that the degree of
$J/\psi$ absorption by produced soft particles is small and cannot
account for the $J/\psi$ data in Pb-Pb collisions.  The anomalous
suppression of $J/\psi$ production in Pb-Pb collisions can be
explained as due to the occurrence of a new phase of strong $J/\psi$
absorption, which sets in when the local energy density exceeds about
3.4 GeV/fm$^3$.  To probe the chemical content of the new phase, we
propose to study the abundance of open-charm mesons and charm hyperons
which depends sensitively on the quark chemical potential.

\end{abstract}

\section{INTRODUCTION}

Following the initial suggestion by Matsui and Satz \cite{Mat86}, the
NA50 experimental observation \cite{Gon96,Lou96,Ger97} of a possible
discontinuity in the absorption of $J/\psi$ in Pb-Pb collisions has
led to a flurry of theoretical activities.  While the present author
and others presented theoretical findings at the Quark Matter Meeting
in 1996 \cite{Won96qm,Kha96qm,Bla96qm,Gav96qm}, other theoretical
studies have since been put forth
\cite{Won97,Cap96,Cas96,Kha96a,Arm96,Sor97}.  A central question is
whether the anomalous suppression in Pb-Pb collisions arises from the
occurrence of a new phase of strong $J/\psi$ absorbing matter
(possibly a quark-gluon plasma), as suggested in
\cite{Won96qm,Kha96qm,Bla96qm,Won97,Kha96a}, or from the absorption by
comovers (produced hadrons), as proposed by
\cite{Gav96qm,Cap96,Cas96}.

To shed light on this question, we shall examine the absorption
mechanisms both qualitatively and quantitatively and study how
different mechanisms can be recognized by their characteristic
signatures.  From these studies, the deviation of the $J/\psi$ data in
Pb-Pb collisions from the results extrapolated from $pA$, O-Cu, O-U,
and S-U collisions suggests that there is a new absorption mechanism
which results in the anomalous suppression of $J/\psi$ production in
Pb-Pb collisions.

While the debate on the theoretical interpretation of the experimental
data continues, it is useful to look for other effects of a
quark-gluon plasma so as to provide means for its search and
identification.  We shall examine a possible signature which allows
one to study the chemical content of the quark-gluon plasma, if it is
ever produced.
\vskip -0.3cm
\section{  PRODUCTION AND ABSORPTION of $J/\psi$ AND $\psi'$}
\vskip -0.1cm
In a nucleon-nucleon collision, $c\bar c$ pairs are produced by the
collision of a parton of one nucleon with a parton of another nucleon.
Among the $c\bar c$ pairs, those $c\bar c$ pairs with large relative
energies will separate into $D\bar D$ pairs and will not be the
subject of our attention.  Only those with low invariant masses will
have a high probability to evolve into precursors of $J/\psi$ and
$\psi'$.  These precursors can be in a color-singlet or color-octet
state \cite{Bod95}.  With respect to a nucleon, a precursor in a
color-octet state has a total cross section of a few tens of
millibarns, which is much greater than that in a color-singlet state
\cite{Dol92,Woncw96b}.  If the precursor is produced as an incoherent
mixture of the two color states, then the survival probability for the
precursor will be the sum of two different exponential survival
probabilities with two characteristic absorption lengths.  In
principle, the color-octet production fraction can be inferred from
the absorption curve in $pA$ collisions as a function of the atomic
number $A$ \cite{Won96b}.  However, the sparsity of the data points
and the uncertainty of the experimental measurements do not allow a
clean separation of the two color fractions.  The analysis of
experimental data suggests that NA3 \cite{Bad83}, NA38 \cite{Bag89},
and E772 \cite{Ald91} data are not inconsistent with the theoretical
picture \cite{Tan96,Ben96} that color-octet and color-singlet
precursors are produced in roughly equal proportions if the produced
color-singlet precursors are pointlike and transparent. However, if
the color-singlet precursors are not transparent but have a cross
section of a few mb, these data do show a definite preference for a
larger fraction of color-singlet precursors \cite{Won96b}. In the
present work, we shall limit our attention to a single effective color
component for our discussions.

A very important property of a precursor is the energy needed to
separate it into a $D\bar D$ pair.  A color-singlet $J/\psi$ particle
needs an energy of 640 MeV to break into a $D\bar D$ pair, while a
$\chi_{1,2}$ needs about 200 MeV, and a $\psi'$ only about 50 MeV.  A
color-octet precursor will need to emit or absorb a soft gluon in the
time scale of nonperturbative QCD to neutralize into a color-singlet
state.  If it is broken up into a $D\bar D$ pair, one of the pair will
be in a color-octet state, and the need to emit or absorb a soft gluon
remains unchanged.  Thus, the average energy required to break a
color-octet $J/\psi$ precursor is still about 640 MeV and a
color-octet $\psi'$ precursor about 50 MeV, but with a large standard
deviation (of about a few hundred MeV), due to the emission or
absorption of the soft gluon.  For simplicity, we shall group $\chi$
particles with the $J/\psi$ because of the large energies required to
break them into $D\bar D$ pairs.

We consider the collision of a projectile nucleus $B$ with a target
nucleus $A$ leading to the production of a $J/\psi$ or $\psi'$
precursor, depicted by the big dot at $P$ in Fig. 1. The precursor
will collide with two groups of particles at different energies: the
projectile and target nucleons and the produced soft particles.  The
dynamics of precursor-particle collisions is best studied in the
nucleon-nucleon center-of-mass frame illustrated in Fig. 1, where
$\gamma$ is the Lorentz contraction factor.  In this frame, the
nucleons still retain a large
fraction of their initial kinetic energies.  As $J/\psi$ and $\psi'$
are produced predominantly at
central rapidity, the precursors can be envisaged as produced nearly
at rest.  Collisions between the nucleons and the precursor take place
at high energies, much higher than the\break 
\vskip -1.3cm \hangafter=-16 \hangindent=-3in 
\noindent
separation thresholds of the
$J/\psi$ or $\psi'$ precursor, and likely lead to the breakup of the
precursor into a $D\bar D$ pair.  Absorption due to collisions with
nucleons can be called the hard component.  The precursor survival
probability, after colliding with projectile and target nucleons, is
$\exp\{ -\sigma(J/\psi$-$N) \rho L\}$ for $J/\psi$ and is $\exp\{
-\sigma(\psi'$-$N) \rho L\}$ for $\psi'$, where $L$ is the sum of the
path length $L_B$ in the projectile and $L_A$ in the target nucleus in
their respective rest frames, $\rho$ is the nucleon density at rest,
and $\sigma(J/\psi$-$N)$ and $\sigma(\psi'$-$N)$ are respectively the
$J/\psi$-$N$ and $\psi'$-$N$ absorption cross section \cite{Ger88}.

\null\vskip 1.0cm \epsfxsize=250pt
\includegraphics{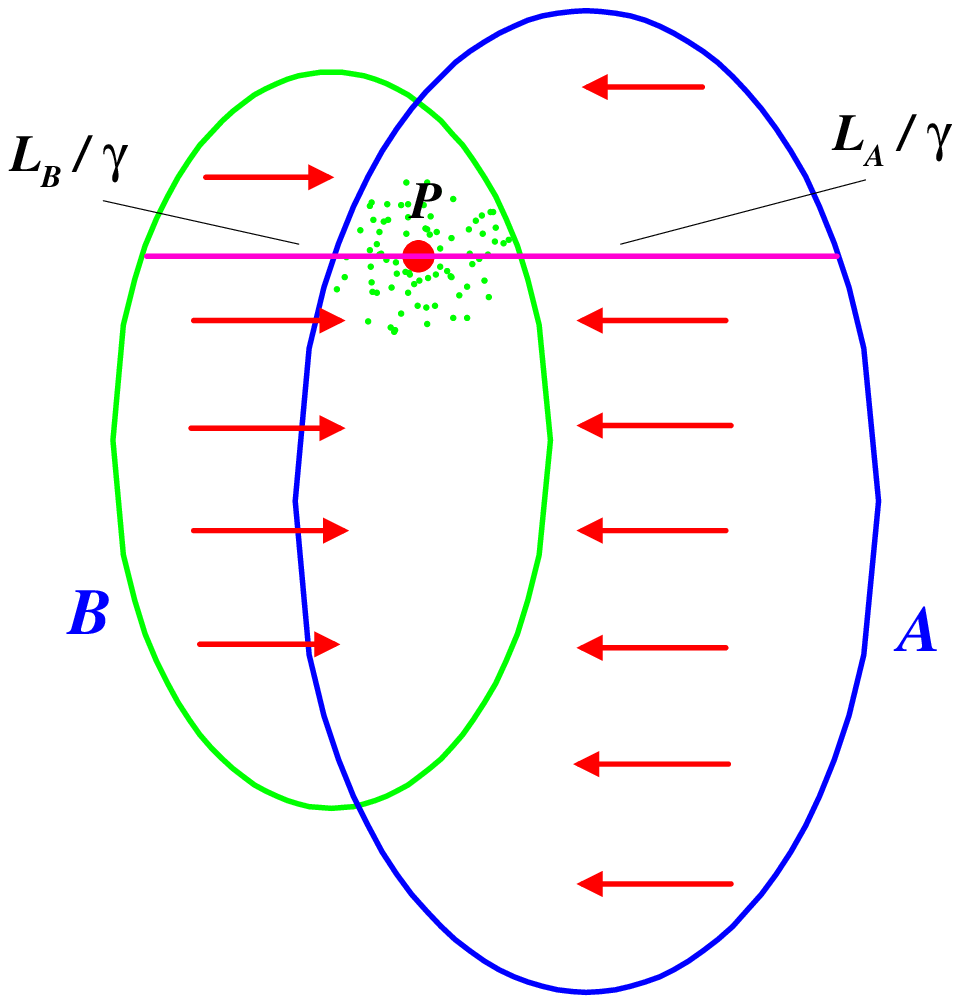}
\vskip -2.8cm
\null\hskip 8.4cm
\begin{minipage}[t]{6.7cm}
\noindent
\bf Fig.1. { \rm
{ The collision of a projectile nucleus $B$ with a target nucleus $A$ in
the nucleon-nucleon center-of-mass\break system.}}
\end{minipage}
\vskip 4truemm
\noindent 

\vskip -1.6cm \hangafter=-3 \hangindent=-3in 
From the space-time picture of the collision process, we
can infer that the hard absorption component is operative for both
$pA$ and\break (nucleus $A$)-(nucleus $B$) collisions.  However, in
nucleus-nucleus collisions, there is an additional absorption
component.  There, a $J/\psi$ or $\psi'$ precursor produced at $P$ in
Fig. 1 finds itself in the middle of fireballs of soft particles
produced by earlier or later nucleon-nucleon collisions centered at
the same spatial location (soft particles depicted by the small dots
around the point $P$ in Fig. 1).  These produced soft particles may
exist in the form of virtual gluons in early stages and hadrons at
later stages \cite{Web84}.  They will collide with the $J/\psi$ or
$\psi'$ precursor and lead to its breakup.  The centers of the
fireballs of produced soft particles are nearly at rest, and the
collisions between the precursor and the soft particles of the
fireballs occur at a kinetic energy which is about twice the fireball
temperature (of a few hundred MeV).  Absorption due to these
collisions can be called the soft component of $J/\psi$ and $\psi'$
absorption.

The survival probability of $J/\psi$ and $\psi'$ due to the absorption
by soft particle collisions is approximately an exponential function
whose exponent is proportional to the (precursor)-(soft particle)
absorption cross section and the density of soft particles in which
the precursor finds itself.  The number of produced soft particles
depends on the number of participants \cite{Sor88}.  The number of
participants in turn is proportional to the longitudinal path length
passing through nuclei $A$ and $B$.  Because of such dependence on
participant numbers and the longitudinal path length, the survival
probability due to soft particle absorption is approximately $\exp\{
-c \sigma({\rm precursor}$-(soft particle))$ L\}$, where $c$ is
approximately a constant (see pages 374-377 of Ref. \cite{Won94}).

From these considerations, we can separate out the two components of
absorption by plotting the logarithm of the survival probability as a
function of the path length.  Following the work of Gerschel and H\"
ufner\cite{Ger88}, the hard component can be separated out by the $pA$
line, and the precursor-nucleon cross section at high energies can be
measured by the slopes of the $pA$ lines.  In addition, the effect of
the soft component can be uncovered by the gap and the change of the
slope going from the $pA$ line to the $AB$ lines.  The magnitude of
the gap and the change of the slope are not independent.  A large gap
is accompanied by a large slope change.  They are related to the
magnitude of the (precursor)-(soft particle) cross section at low
energies.  We expect that because the collision energy in the soft
absorption component is small compared to the separation threshold of
640 MeV for $J/\psi$ but large compared to the separation threshold of
50 MeV for $\psi'$, the absorption behavior for $J/\psi$ and $\psi'$
can be quite different for the soft component.

\section{ COMPARISON WITH EXPERIMENTAL DATA}

We can use the qualitative signatures discussed in the last section to
examine the experimental $J/\psi$ and $\psi'$ data in two different
ways.  We look first at experimental $J/\psi$ and $\psi'$ cross
sections for $AB$ collisions when all impact parameters are summed
over.  If there were no absorption, these cross sections would be
proportional to the product $AB$.  The quantities ${\cal
B}(J/\psi)\sigma(J/\psi)/AB$ and ${\cal B}(\psi')\sigma(\psi')/AB$
(where ${\cal B}(J/\psi)$ and ${\cal B}(\psi')$ are branching
constants) give respectively the $J/\psi$ and $\psi'$ survival
probabilities, multiplied by constants.  In Figs. 1$a$ and $1b$, these
quantities are presented in a semi-log plot as a function of $\eta=
A^{1/3}(A-1)/A + B^{1/3}(B-1)/B $, which is proportional to the path
length averaged over all impact parameters \cite{Won94}. Here, the
quantities $A^{1/3}$ and $B^{1/3}$ are proportional to the radii of
the two nuclei, and the factors $(A-1)/A$ and $(B-1)/B$ are included
to take into account the finiteness of the number of nucleons.

\vskip -0.5cm
\epsfxsize=400pt
\includegraphics{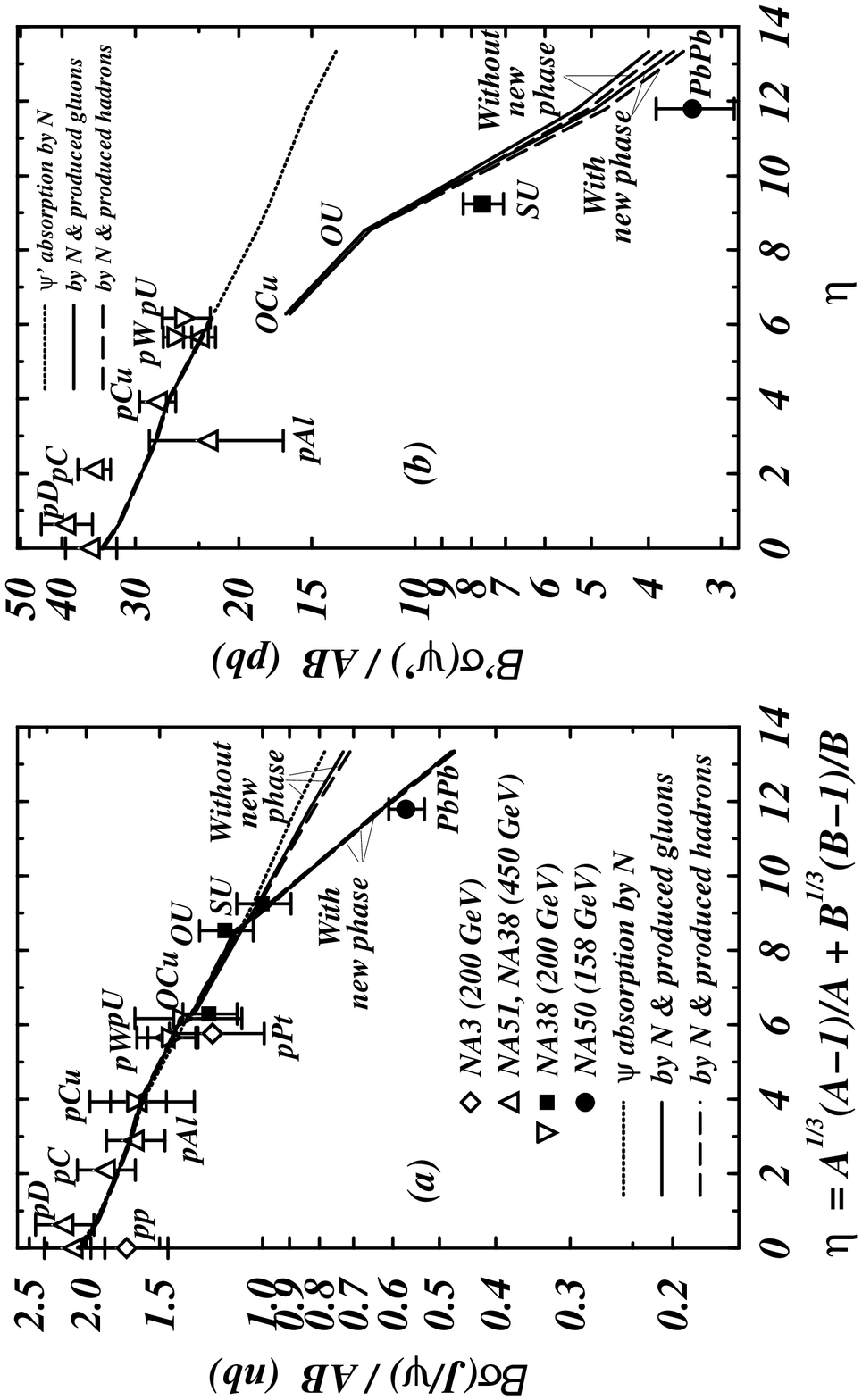}
\vskip 10cm
\begin{minipage}[t]{6in}
\noindent \bf Fig.2.  \rm
{ ($a$) \protect${\cal B}\sigma_{J/\psi}^{AB}/AB\protect$ and
($b$) \protect${\cal B}'\sigma_{\psi'}^{AB}/AB\protect$ as a function
of $\eta$.  Data are from NA3 \protect\cite{Bad83}, NA51
\protect\cite{Bal94}, NA38 \protect\cite{Bag89,Lou95,Bag95}, and NA50
\protect\cite{Gon96,Lou96}.  }
\end{minipage}
\vskip 4truemm
\noindent 

We look next at a different set of data where one uses the information
on the transverse energy to select the effective impact
parameter. This allows one to study the survival probability as a
function of the impact parameter and its associated average path
length $L$.  The $J/\psi$ and $\psi'$ production cross section in the
absence of absorption is proportional to the Drell-Yan cross section.
Because the Drell-Yan cross section does not suffer much absorption in
nuclear and hadron environments, the ratios of ${\cal B}(J/\psi)
\sigma(J/\psi)$ and ${\cal B}(\psi') \sigma(\psi')$ to the Drell-Yan
cross section are good representations of the $J/\psi$ and $\psi'$
survival probabilities, multiplied by constants.  These ratios as a
function of the path length $L$ are presented for $J/\psi$ in
Fig. 2$a$ and for $\psi'$ in Fig. 2$b$.

In Figs. 1 and 2, we show the $pA$ data as open points, the $AB$ data
for O-Cu, O-U and S-U collisions as solid squares, and the Pb-Pb data
as solid circles.  From the data in Fig. 1, we note that the $pA$ data
give approximately straight lines conforming to the signature of the
hard absorption component.  The slopes of the $pA$ lines for $J/\psi$
and $\psi'$ are nearly the same.  Therefore for the hard component of
the absorption process, the rate of absorption and the associated
absorption cross sections by collisions with nucleons at high energies
are approximately the same for $J/\psi$ and $\psi'$.  It is worth
noting that the approximate equality is supported by other
experimental data of $\sigma_{\rm tot}(\psi'$-$N)/\sigma_{\rm
tot}(\psi$-$N)\approx 0.75 \pm 0.15$ inferred from the photoproduction
of $J/\psi$ and $\psi'$ \cite{Bri83}.

\null\vskip -1.0cm
\epsfxsize=400pt
\includegraphics{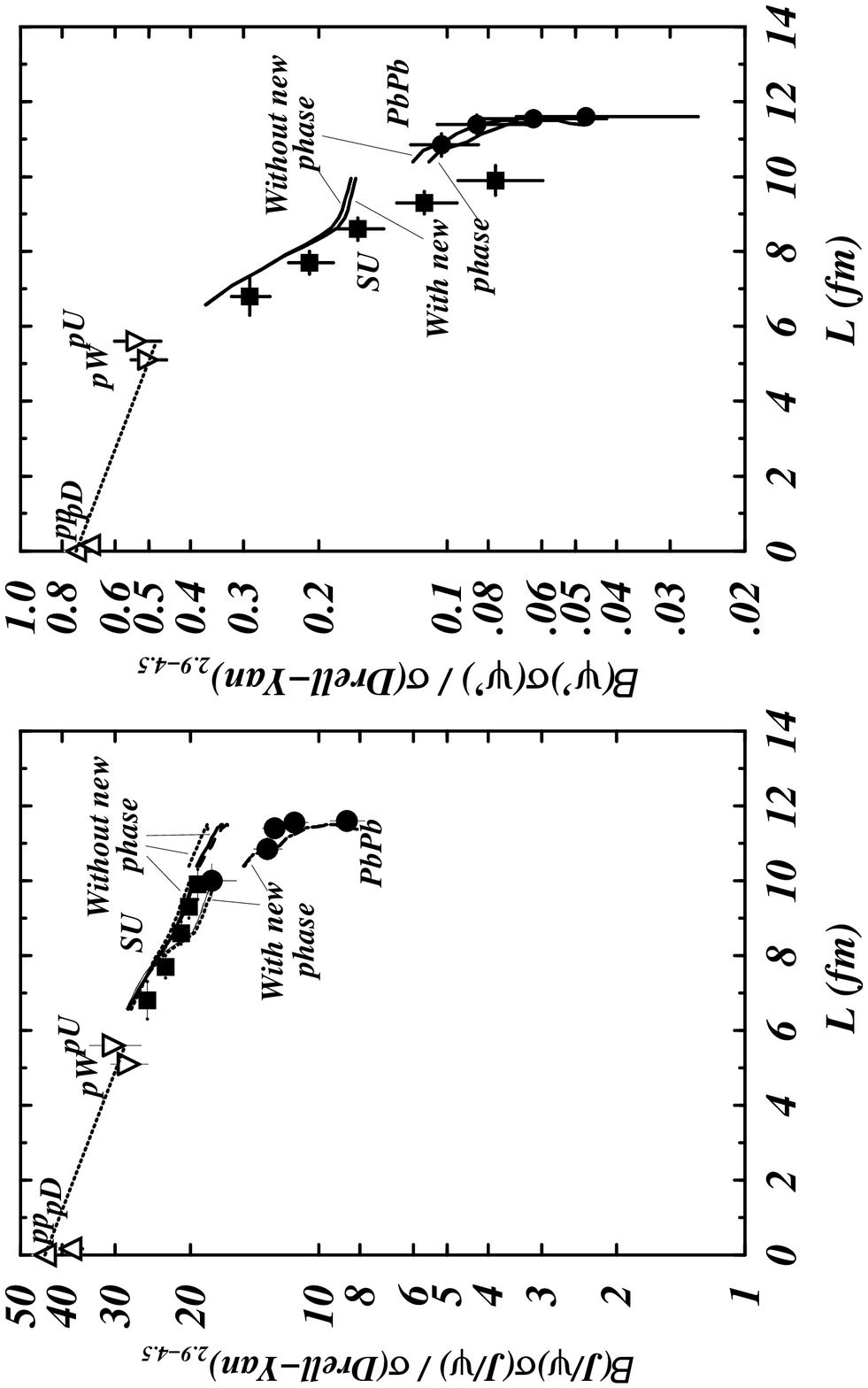}
\null\vskip 9.2cm
\begin{minipage}[t]{6in}
\noindent \bf Fig.3.  \rm
{ ($a$) \protect${\cal
B}\sigma_{J/\psi}^{AB}/AB\protect$ and ($b$) \protect${\cal
B}'\sigma_{\psi'}^{AB}/AB\protect$ as a function of the path length
$L$. Data are from NA51
\protect\cite{Bal94}, NA38 \protect\cite{Bag89,Lou95,Bag95}, and NA50
\protect\cite{Gon96,Lou96}. 
}
\end{minipage}
\vskip 4truemm
\noindent

Theoretically, the approximate equality of $\sigma_{\rm
tot}(\psi'$-$N) \sim \sigma_{\rm tot}(\psi$-$N)$ can be explained by
the Glauber picture of hadron-hadron collisions in terms of
quark-quark collisions as a generalization of the additive quark model
at high energies, for which the absorption cross section for
$\psi$-$N$ collisions and $\psi'$-$N$ collisions can be shown to be
insensitive to the separation between $c$ and $\bar c$ \cite{Won96a}.
The approximate equality leads immediately to the experimental
observation of the independence of $\psi'/\psi$ with mass numbers in
$p$-A collisions \cite{Lou95,Bag95}.  Other, different theoretical
models interpret the approximate equality as arising from the same
separation between the color-octet $(c\bar c)_8$ and a gluon in the
hybrid $(c\bar c)_8$-$g$ for $J/\psi$ and $\psi'$ production
\cite{Kha96}, or a coherent $J/\psi$-$\psi'$ admixture
\cite{Huf96,Woncw96b}.  As hadron-hadron cross sections involve
nonperturbative aspects of QCD dynamics, much more work remains to be
done to clarify the origin of this approximate equality.

To study the soft component of absorption, one looks for a gap and a
change of the slope in going from the $pA$ line to the $AB$ line.
From the $\psi'$ data in Fig. 1$b$, one can discern the presence of a
gap in going from the $pA$ line to the S-U data point.  The $\psi'$
data in Fig. 2$b$ show a big gap and a large change of the slope in
going from the $pA$ line to the $AB$ line, conforming to the signature
of the soft component.  Figs.\ 1$b$ and 2$b$ indicate that for $\psi'$
production, the absorption due to the soft component is large.

The above analysis of the $\psi'$ data provides us with a clear
signature of the soft absorption component.  This signature can be
used to identify the soft component in other production processes.
Upon searching for the signature of the soft absorption component in
$J/\psi$ production data for $pA$, O-Cu, O-S, and S-U collisions in
Figs. 1$a$ and 2$a$, one finds that there is almost no gap and no
change of the slope in going from the $pA$ line to the $AB$ line.  One
concludes that the absorption of $J/\psi$ by soft particles, as
revealed by the data of $pA$, O-Cu, O-U, and S-U, is small.

The difference in the behavior of the $J/\psi$ precursor and $\psi'$
precursor with regard to absorption by soft particles can be easily
understood as due to their difference in the threshold energies
mentioned above.  The average relative kinetic energies between the
soft particles and the precursor are much below the $J/\psi$ threshold
but much above the $\psi'$ threshold, and thus the rate of absorption
of $J/\psi$ by soft particle collisions is small compared with the
rate of absorption of $\psi'$ by soft particle collisions.
 
When one extends one's consideration to Pb-Pb collisions, one finds
that the $AB$ line of O-Cu, O-U, and S-U in Figs.\ 1$a$ and 2$a$ are
much above the Pb-Pb data points.  This indicates that the degree of
$J/\psi$ absorption by soft particles, as revealed by the data of
O-Cu, O-U, and S-U collisions, cannot explain the Pb-Pb data, and a
new phase of strong $J/\psi$ absorption in Pb-Pb collisions is
suggested.

\section{MICROSCOPIC ABSORPTION MODEL OF $J/\psi$ and $\psi'$ ABSORPTION}

We can be more quantitative to study this departure of Pb-Pb data by
using the microscopic absorption model of \cite{Won96a}.
In this model, each nucleon-nucleon collision is a possible source of
$J/\psi$ and $\psi'$ precursors.  It is also the source of a fireball
of soft particles which can absorb $J/\psi$ and $\psi'$ precursors
produced by other nucleon-nucleon collisions.  One follows the
space-time trajectories of precursors, baryons and the centers of the
fireballs of soft particles.  Absorption occurs when the space-time
trajectories of the precursors cross those of other particles.  Using
a row-on-row picture in the center-of-mass system and assuming
straight-line space-time trajectories, we obtain the differential
cross section for $J/\psi$ production in an $AB$ collision as
\cite{Won96a}
\begin{eqnarray}
\label{eq:fin}
{ d \sigma_{{}_{J/\psi}}^{{}^{AB}} ({\bf b}) \over \sigma_{{}_{J/\psi
}}^{{}^{NN}}~d{\bf b} } &=& \!\! \int \!\! { d{ {\bf b}}_{{}_A} \over
\sigma_{\rm abs}^2(J/\psi- N) } \biggl \{ 1 -\biggl [ 1-
T_{{}_A}({\bf b}_{{}_A}) \sigma_{\rm abs}(J/\psi-N) \biggr ] ^A \biggr
\} \nonumber\\
& &~~~~~~~\times \biggl \{ 1 -\biggl [ 1- T_{{}_B}({\bf b}-{\bf b}_{{}_A})
\sigma_{\rm abs}(J/\psi- N) \biggr ] ^B \biggr \} F({\bf b}_A, {\bf b})\,,
\end{eqnarray}
where $T_A(\bf{b}_A)$ is the thickness function of nucleus $A$, and
$F({\bf b}_A, {\bf b})$ is the survival probability due to soft
particle collisions.  To calculate $F({\bf b}_A, {\bf b})$, we sample
the target tranverse coordinate ${\bf b}_A$ for a fixed impact
parameter ${\bf b}$ in a row with the nucleon-nucleon inelastic cross
section $\sigma_{in}^{NN}$.  In this row, $BT_B({\bf b}-{\bf
b}_A)\sigma_{in}$ projectile nucleons will collide with $AT_A({\bf
b}_A)\sigma_{in}$ target nucleons.  We construct the space-time
trajectories of these nucleons to locate the position of their
nucleon-nucleon collisions.  These collisions are the sources of
$J/\psi$ and $\psi'$ precursors and the origins of the fireballs of
produced particles.  For each precursor source from the collision $j$
and each absorbing fireball from the collision $i$ at the same spatial
location, we determine the time when the precursor source coexists
with the absorbers as virtual gluons $t_{ij}^g$ or as produced hadrons
$t_{ij}^h$.  The survival probability due to this combination of
precursor source and absorber is then $\exp \{-k_{\psi g} t_{ij}^g
-k_{\psi h} t_{ij}^h \}$, where the rate constant $k_{\psi m}$ for
$m=g, h$ is the product of $\sigma_{\rm abs}(J/\psi$-$ m)$, the
average relative velocity $v_m$, and the average number density
$\rho_m$ per $NN$ collision.  When we include all possible precursor
sources and absorbers, $F({\bf b}_A,{\bf b})$ becomes
\begin{eqnarray}
\label{eq:fb}
F({\bf b}_A, {\bf b}) \!= \!\sum_{n=1}^{N_<}  \!{ a(n)\over {N}_> {N}_<} 
\!\sum_{j=1}^n
\!\exp\{- \theta \!\!\!\! \sum_{i=1, i\ne j}^n 
\!\!\!( k_{\psi g} t_{ij}^g +
k_{ \psi h} t_{ij}^h ) \} \,,
\end{eqnarray}
where $N_>({\bf b}_A)$ and $N_<({\bf b}_A)$ are the greater and the
lesser of the (rounded-off) nucleon numbers $AT_A({\bf
b}_A)\sigma_{in}$ and $BT_B({\bf b}-{\bf b}_A)\sigma_{in}$, $a(n)=2
{\rm~~for~~} n=1,2,...,N_<-1 $, and $ a(N_<)=N_>-N_<+1$.  The function
$ \theta$ is zero if $A=1$ or $B=1$ and is 1 otherwise.  The survival
probability $F$ can be determined from plausible $c\bar c$, $g$, $h$
production time $t_{c\bar c}$, $t_g$, $t_h$, and the freezeout time
$t_f$ \cite{Won96a}. The cross section for $\psi'$ production can be
obtained from the above equations by changing $J/\psi$ into $\psi'$.

We use this microscopic absorption model to study the experimental
data (see also \cite{Won96qm,Won97}). Consider first the $J/\psi$ data
in $pA$, O-Cu, O-U and S-U collisions.  If one assumes that there is
no soft particle absorption, the results with the least $\chi^2$ are
obtained with $\sigma_{\rm abs}(J/\psi$-$N)=6.94$ mb, shown as dotted
curves in Fig. 1$a$ and 2$b$.  If one assumes nucleons and virtual
gluons as absorbers, the least $\chi^2$ fits shown as the solid curves
marked by ``without new phase'' in Figs. 1$a$ and 2$a$ are obtained
with $\sigma_{\rm abs}(J/\psi$-$N)=6.36$ mb and $k_{\psi g}=0.049$
c/fm (which corresponds to a $J/\psi$-gluon cross section of about 0.3
mb).  Alternatively, if one assumes nucleons and produced hadrons as
absorbers, the least $\chi^2$ fits shown as the dashed curve marked by
``without new phase'' are obtained with $\sigma_{\rm
abs}(J/\psi$-$N)=6.36$ mb and $k_{\psi g}=0.096$ c/fm (which
corresponds to a $J/\psi$-hadron cross section of about 0.68 mb).  The
results in Figs. 1$a$ and 2$a$ indicate that the soft component of
$J/\psi$ absorption as revealed by O-Cu, O-U, and S-U collisions is
small, and the extrapolated results from any one of these three
descriptions are much greater than the Pb-Pb data points.  The Pb-Pb
data cannot be explained by the absorption due to collisions with
nucleons and soft particles.

We next examine the $\psi'$ data in Figs. 1$b$ and 2$b$.  The
theoretical results with no soft particle absorption are given by the
dotted curves, which fit the $pA$ data, but are much too large for the
S-U data.  Theoretical results calculated with
$\sigma(\psi'$-$N)=6.36$ mb, and $k_{\psi g}=3$ c/fm (which
corresponds to a $\psi'$-gluon cross section of about 20 mb for
virtual gluon as absorbers) are shown as the solid curve marked by
``without new phase'' in Fig. 1$b$ and as the upper solid curves in
Fig. 2$b$.  Theoretical results calculated with $\sigma(\psi'-N)=6.36$
mb, and $k_{\psi h}=3$ c/fm (which corresponds to a $\psi'$-hadron
cross section of about 20 mb for produced hadrons as absorbers) are
shown as the dashed curve marked by ``without new phase'' in
Fig. 1$b$ and coincide with the upper solid curves in Fig. 2$b$.
These theoretical results show that soft particle absorption leads to
a gap and a change of the slope when one goes from the $pA$ line to
the $AB$ line.  A large soft component is needed to explain the
$\psi'$ data in S-U collisions.  The flattening of the theoretical
ratio of $B(\psi')\sigma(\psi')/\sigma({\rm Drell-Yan})$ as a function
of $L$ for S-U collisions in Fig. 2$b$ arises because the distribution
of soft particle densities in the central region of $\psi'$ production
is insensitive to the impact parameter for small impact parameters
when the masses of the two colliding nuclei are very different.
Nuclear deformation may play a role in leading to a greater density of
soft particles for collisions with the largest transverse energies and
may lead to the discrepancy of the theoretical results with
experimental data for those S-U collisions with the largest transverse
energies (or the largest apparent path lengths in Fig. 2$b$).  The
comparison of S-Pb and S-U collisions will shed light on the
deformation effect.

\section{NEW PHASE OF $J/\psi$ ABSORPTION}

The deviation of the $J/\psi$ data in Pb-Pb collisions from the
conventional theoretical extrapolations in $p$-A, O-A, and S-U
collisions suggests that there is a transition to a new phase of
strong absorption, which sets in when the local energy density exceeds
a certain threshold.  We can extend the absorption model to describe
this transition.  The energy density at a particular spatial point at
a given time is approximately proportional to the number of
nucleon-nucleon collisions which has taken place at that spatial point
up to that time.  We use the row-on-row picture as before, and
postulate that soft particles make a transition to a new phase with
stronger $J/\psi$ absorption characteristics at a spatial point if
there have been $N_c$ or more baryon-baryon collisions at that spatial
point.  The quantity $k_{\psi g} t_{ij}^g + k_{\psi h} t_{ij}^h$ in
Eq.\ (\ref{eq:fb}) becomes $k_{\psi g} t_{ij}^g + k_{\psi h} t_{ij}^h
+ k_{\psi x} t_{ij}^x$.  Here, the new rate constant $k_{\psi x}$
describes the rate of absorption of $J/\psi$ by the produced soft
matter absorber in the new phase. Also, the quantity $t_{ij}^x$ is the
time for a $J/\psi$ produced in collision $j$ to coexist at the same
spatial location with the absorbing soft particles produced in
collision $i$ in the form of the new phase, before hadronization takes
place.  Baryons passing through the spatial region of the new phase
may also become deconfined to alter their $J/\psi$ absorption
characteristics.  Accordingly we also vary the effective absorption
cross section, $\sigma_{\rm abs}^x(\psi N)$, for $\psi$-$N$
interactions in the row in which there is a transition to the new
phase, while the absorption cross section $\sigma_{\rm abs}(\psi N)$
remains unchanged in other rows where there is no transition.  As
shown on the curves marked by ``with new phase'' in Figs.\ 1$a$ and
2$a$, model results assuming a new phase give good agreement with
${\cal B}\sigma_{J/\psi}^{AB}/AB$ data including the Pb-Pb data point,
with the parameters $N_c=4$, $k_{\psi x}=1$ c/fm.  The rate constant
$k_{\psi x}$ for this new phase is much greater than the corresponding
rate constants $k_{\psi g}$ or $k_{\psi h}$.

We can study the $\psi'$ data to see how the presence of the new phase
will affect $\psi'$ production.  Theoretical results obtained by
assuming the new phase are shown as the curves marked by ``with new
phase'' in Fig. 1$b$ and the lower solid curves in Fig. 2$b$.  They
indicate that as far as $\psi'$ suppression is concerned, the $\psi'$
particles are so strongly absorbed by collisions with soft particles
that the presence of an additional source of absorption leads only to
a very small additional absorption.  Seen in this light, $J/\psi$ is a
better probe for a new phase of absorption than $\psi'$ because of its
large threshold value which does not allow it to be destroyed in great
proportion by soft particles.

We have seen that the anomalous suppression of $J/\psi$ in Pb-Pb
collisions can be explained by models in which a new phase of strong
absorption sets in when the number of baryon-baryon collisions at a
local point exceeds or is equal to $N_c=$4.  We can inquire about the
approximate threshold energy density $\epsilon_c$ which corresponds to
the onset of the new phase.  Evaluated in the nucleon-nucleon
center-of-mass system, the energy density at the spatial point, which
has had $N_c$ prior nucleon-nucleon collisions, is approximately
\begin{eqnarray}
\epsilon_c= N_c {dn \over dy} {m_t \over \sigma_{in} d/\gamma}
\end{eqnarray}
where $dn/dy\sim 1.9$ \cite{Tho77} is the particle multiplicity per
unit of rapidity at $y_{{}_{CM}}=0$ for an $NN$ fixed-target collision
at 158A GeV, $m_t\sim 0.35$ GeV is the transverse mass of a produced
pion, $d \sim 2.46$ fm is the internucleon spacing, and
$\gamma=\sqrt{s}/2m_{\rm nucleon}=9.2$ is the Lorentz contraction
factor.  For $N_c=4$ we find $\epsilon_c \sim 3.4$ GeV/fm$^3$, which
is close to the quark-gluon plasma energy density, $\epsilon_c \sim
4.2$ GeV/fm$^3$, calculated from the lattice gauge theory result of
$\epsilon_c/T_c^4 \sim 20$ \cite{Blu95} with $T_c \sim 0.2$ GeV.
Therefore, it is interesting to speculate whether the new phase of
strong absorption may be the quark-gluon plasma.  In an equilibrated
or non-equilibrated quark-gluon plasma, the screening of the $c$ and
$\bar c$ quarks by deconfined quarks and deconfined gluons will weaken
the interaction between $c$ and $\bar c$ and will enhance the breakup
probability of a quasi-bound $(c \bar c)$ system.  

\section{OTHER SIGNATURES OF THE QUARK-GLUON PLASMA}

The above analysis shows that some domains of matter produced in Pb-Pb
collisions may be in a new phase of strong $J/\psi$ absorption.
Whether this new absorbing matter is the quark-gluon plasma will
require collaborative evidence.  If the quark-gluon plasma can
suppress $J/\psi$ production, it will also have other detectable
consequences.  In particular, it will affect the abundance of the $u,
d$, and $s$ quarks and antiquarks which may lead to a definite pattern
of the abundance of the $D$ mesons and charm hyperons.  The abundance
of open-charm mesons and charm hyperons may be used to probe and
identify the chemical content of the quark-gluon plasma.

\null\vskip 8.4cm 
\epsfxsize=250pt
\includegraphics{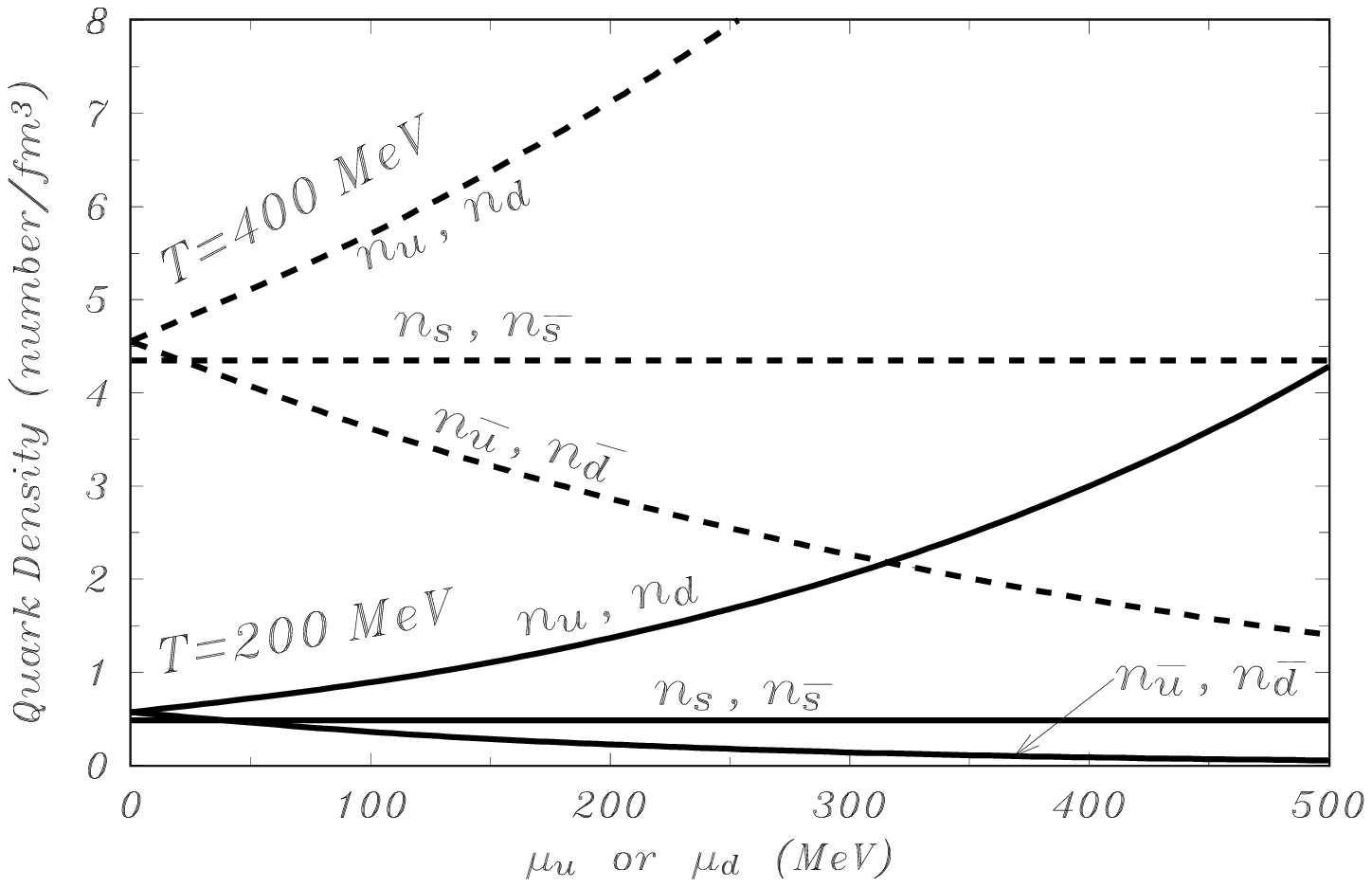}
\vskip -1.9cm
\null\hskip 0.7in\begin{minipage}[t]{4.5in}
\noindent \bf Fig.4.  \rm { Densities of various types of quarks and
antiquarks as a function of the up and down quark chemical potential.
}
\end{minipage}
\vskip 2truemm
\noindent 

To illustrate the salient features of the physics, we can consider a
free quark gas with three favors at thermal and chemical equilibrium.
The quark and antiquark densities as a function of the chemical
potential of $u$ and $d$ quarks are given in Fig. 4 (from Fig. 18.2 of
Ref. \cite{Won94}).  For a quark-gluon plasma with no net baryon
content, as would be expected in the ``transparent region'' in very
energetic heavy-ion collisions when the baryons are not stopped but
proceed forward after collision, the densities of all quark species
are approximately equal.  On the other hand, for a quark-gluon plasma
with a large net baryon content, such as would be expected in the
``stopping region'' of high-energy heavy-ion collisions, the densities
of $u$ and $d$ quarks are much greater than the densities of $s$ and
$\bar s$ quarks, which are in turn greater than the densities of $\bar
u$ and $\bar d$ quarks.

Consider now the production of $c \bar c$ pairs in nucleon-nucleon
collisions when there is the occurrence of the quark-gluon plasma.  We
focus our attention to those $c\bar c$ pairs which will not form bound
$c\bar c$ states, either because of their large initial relative
kinetic energies or eventual breakup by interaction with other particles.
These $c$ and $\bar c$ particles will coalesce with light quarks and
antiquarks to hadronize when the temperature drops below the
transition temperature.  The abundance of the different types of $D$
and $\bar D$ mesons will depend on the densities of the different
kinds of quarks present in the quark-gluon plasma.

A $\bar c$ can coalesce with $u $, $d$, and $s$ to form $\bar D^0
(\bar c u)$, $D^- (\bar c d)$, and $D_s^- (\bar c s)$ respectively,
while a $c$ can coalesce with $\bar s$, $\bar u$, and $\bar u$ to
form $D_s^+(c \bar s)$, $D^+(c \bar d)$, and $D^0(c \bar u)$
respectively.  Thus, if the chemical potential is zero, the abundances
of the six $D$ mesons will be approximately equal:
\begin{eqnarray}
\bar D^0 (\bar c u) \sim D^- (\bar c d) \sim D_s^- (\bar c s)
\sim D_s^+(c \bar s) \sim D^+(c \bar d) \sim D^0(c \bar u).
\end{eqnarray}

On the other hand, for a quark-gluon plasma with a large net baryon
content having $\mu_u \sim \mu_d$,  the quark densities are related by
$n_u \sim n_d  >> n_s$ and
\begin{eqnarray}
\label{eq:uds}
\bar D^0 (\bar c u) \sim D^- (\bar c d) >> D_s^- (\bar c s).
\end{eqnarray}
For
this case, we have similarly $ n_s = n_{\bar s} >> n_{\bar d} \sim
n_{\bar u} $ and
\begin{eqnarray}
\label{eq:bar}
D_s^- (\bar c s), D_s^+(c \bar s) >> D^+(c \bar d) \sim D^0(c \bar
u).
\end{eqnarray}

As a concrete example, we can study the case of $\mu_u=\mu_d=400$ MeV
at $T=200$ MeV for which the $u$ and $d$ quark densities are about 6
times that of the $s$ and $\bar s$ quarks, which are in turn
approximately 6 times the densities of $\bar u$ and $\bar d$ antiquarks.
In that case, Eqs. (\ref{eq:uds}) and (\ref{eq:bar})
give

\begin{eqnarray}
\bar D^0 (\bar c u) : D^- (\bar c d) : D_s^- (\bar c s)
\approx ~6~:~6~:~1,
\end{eqnarray}
\begin{eqnarray}
{\rm and~~~~}D_s^- (\bar c s) : D_s^+(c \bar s) : D^+(c \bar d) : D^0(c \bar
u) \approx ~6~:~6~:~1~:~1.
\end{eqnarray}

For a chemical potential as large as $\mu_u=\mu_d=400$ MeV, it is more
likely to find two quarks of $u$ and $d$ type than a $\bar u$, $\bar
d$ or an $\bar s$ antiquark.  It will be more likely for the $c$ quark
to combine with $u$ and $d$ quarks to form charm hyperons
$\Lambda_c^+(udc), \Sigma_c^{++}(uuc), \Sigma_c^+(udc),$ and
$\Sigma_c^0(ddc)$ than forming an open charm meson.  The production of
these charm hyperons with $c$ quarks will be enhanced.

It is expected that while the primary abundance will be related to
the chemical potential of the quark-gluon plasma, the final abundance
of the $D$ mesons will be subject to small modifications by
final-state interactions.  In particular, reactions such as $D^{\pm} +
K^{\pm} \rightarrow D_s^{\pm} +\pi^{\pm}$ are exothermic and take
place without additional energy.  They will shift the abundances of
$D^{\pm}$ and $D_s^{\pm}$ and need to be taken into account.  It
will be both an experimental and theoretical challenge to use the
abundance of the $D$ mesons and charm hyperons to probe the chemical
content of the quark-gluon plasma.

\section{CONCLUSIONS AND DISCUSSIONS}

$J/\psi$ and $\psi'$ precursors produced in high-energy heavy-ion
collisions are absorbed by their collisions with nucleons and produced
soft particles, leading to two distinct absorption mechanisms.  These
mechanisms need to be well understood if one wants to extract
information on $J/\psi$ and $\psi'$ absorption by the presence of the
quark-gluon plasma.  The absorption by nucleons occurs at high kinetic
energies between the precursor and the nucleon and constitutes the
hard component of absorption.  It is operative in $pA$ and $AB$
collisions.  The absorption by soft particles occurs at low relative
energies, and constitutes the soft component of absorption.  It occurs
mainly in $AB$ collisions.

The signature for the hard component is an approximately straight line
in the semi-log plot of the survival probability for $pA$ collisions
as a function of the path length or $\eta= A^{1/3}(A-1)/A +
B^{1/3}(B-1)/B $.  The slope of the line gives the precursor-nucleon 
absorption cross section.

The signature for the soft component consists of a gap and a change of
the slope from the $pA$ line to the $AB$ line in the semi-log plot of
the survival probability as a function of the path length or $\eta$.
The greater the gap, the greater is the change of the slope, and vice
versa.  Application of these signatures to examine the $J/\psi$ and
$\psi'$ data indicates that the degree of absorption by soft particles
on $J/\psi$ production, as revealed by the O-Cu, O-U, and S-U data, is
small.  The absorption by soft particles on $\psi'$ production is
however quite large.

A microscopic absorption model which takes care of all precursor
sources and absorbers is used to examine these two mechanisms.  The
microscopic model results support the above qualitative descriptions.

When one extends one's consideration to Pb-Pb collisions, one finds
that the degree of $J/\psi$ absorption by soft particles as
constrained by the data of O-Cu, O-U, and S-U collisions cannot
explain the Pb-Pb data, and a new phase of strong $J/\psi$ absorption
in Pb-Pb collisions is suggested.  The anomalous suppression of
$J/\psi$ production in Pb-Pb collisions can be explained as due to the
occurrence of a new phase of strong $J/\psi$ absorption, which sets in
when the number of nucleon-nucleon collisions at a spatial point
exceeds about 4 and corresponds to a local energy density of about 3.4
GeV/fm$^3$.

In order to demonstrate that the anomalous suppression in Pb-Pb
collisions arises from the occurrence of the quark-gluon plasma, it is
necessary to obtain further collaborative experimental evidence.  One
can use the abundance of D-mesons and charm hyperons to probe the
chemical content of the new absorbing phase, so as to infer the
chemical state of the system.  Much future work remains to be done to
identify the quark-gluon plasma if it is ever produced in high-energy
Pb-Pb collisions.

\end{document}